\documentclass[twocolumn,showpacs,floats,floatfix,superscriptaddress,aps,pra]{revtex4-1}
\usepackage{amsfonts}
\usepackage{amssymb}
\usepackage{amsmath}
\usepackage{calc}
\usepackage{graphicx}
\usepackage{bm}

\usepackage[normalem]{ulem}
\usepackage{amsmath,amssymb}
\usepackage{multirow}
\usepackage{xcolor,soul}
\usepackage{srcltx}
\usepackage{hyperref,graphicx}

\def\be{ \begin{equation}}
\def\ee{ \end{equation}}
\def\bea{ \begin{eqnarray}}
\def\eea{ \end{eqnarray}}
\def\bse{ \begin{subequations}}
\def\ese{ \end{subequations}}
\def\bc{ \begin{center}}
\def\ec{ \end{center}}

\begin{document}

\author{Stefano Longhi$^{*}$} 
\affiliation{Dipartimento di Fisica, Politecnico di Milano, Piazza L. da Vinci 32, I-20133 Milano, Italy}
\affiliation{IFISC (UIB-CSIC), Instituto de Fisica Interdisciplinar y Sistemas Complejos, E-07122 Palma de Mallorca, Spain}
\email{stefano.longhi@polimi.it}

\title{Phase transitions in non-Hermitian superlattices}
  \normalsize

%\date{.}

%
\bigskip
\begin{abstract}
\noindent  
 We investigate the energy spectral phase transitions arising in one-dimensional superlattices under an imaginary gauge field and possessing $M$ sites in each unit cell in the large $M$ limit. It is shown that in models displaying nearly flat bands a smooth phase transition, from quasi entirely real to complex energies, can be observed as the imaginary gauge field is increased, and that the phase transition becomes sharper and sharper (exact) as $M$ is increased. In this limiting case, for superlattices with random or incommensurate disorder the spectral phase transition corresponds to a localization-delocalization transition of the eigenfunctions within each unit cell, dubbed non-Hermitian delocalization transition and originally predicted by Hatano and Nelson. However, it is shown here that in superlattices without disorder a spectral phase transition can be observed as well, which does not correspond to a non-Hermitian delocalization phase transition. The predicted phenomena could be observed in non-Hermitian photonic quantum walks, where synthetic superlattices with controllable $M$ and imaginary gauge fields can be realized with existing experimental apparatus.
 \end{abstract}

\maketitle
   
\section{Introduction}
The interplay between non-Hermiticity, disorder and topology is a rapidly evolving and timely area of research encompassing different fields  of physics, from condensed matter physics and cold atoms to classical systems, such as  photonic, acoustic, electric and mechanical systems \cite{h1,h2,h3,h3b,h3c,h3d,h3e,h3f,h3g,h4,h5,h6,h7,h8,h9,h10,h11,h12,h13,h14,h15,h16,h17,h17b,h18,h19,h20,h21,h22,h23,h24,h25,h26,h27,h28,h29,h30,h31,h32,h32b,h33,h34,h35,h36,h37,h38,h39,h40,h41,h42,h43,h44,h45,
h46,h47,h48,h49,h50,h51,h52,h53,h54,h55,h56}. The great interest in such a kind of research is motivated, on the one hand, by the richer phenomenology and topology
that systems described by effective non-Hermitian (NH) Hamiltonians display as compared to conservative (Hermitian) systems (see e.g. \cite{U1,U2,U3,U4,U5,U6,U7,U8,U9,U10,U11,U12,U13} and references therein); on the other hand, the recent experimental advances in engineering dissipation in synthetic matter, notably using photonic and cold-atom lattices, offer a great flexibility in controlling and observing a plethora of non-Hermitian phenomena \cite{e1a,e1b,e1c,e1d,e2,e3,e4,e5,e6,e7,e8,e9,e10}.\\
A paradigmatic NH model displaying remarkable physical properties is provided by the Hatano-Nelson model \cite{h1,h2}, which was originally introduced to describe Anderson 
localization in a one-dimensional tight-binding lattice with uncorrelated on-site potential disorder and with asymmetric hopping amplitudes, arising from a superimposed imaginary (rather than real) magnetic flux. Hatano and Nelson showed that, under periodic boundary conditions, the imaginary gauge  
 field can prevent Anderson localization,
with the appearance of a mobility interval at the center
of the band. Interestingly, such a non-Hermitian delocalization transition is associated to a spectral phase transition: while localized eigenstates of the Hamiltonian correspond to real eigenenergies, delocalized eigenstates emerging near the center of the band correspond to complex eigenenergies. The non-Hermitian delocalization transition has been subsequently reconsidered by
several authors \cite{h3,h3b,h3c,h3d,h3e,h3f,h3g}, and found potential relevance for chiral robust transport in photonic systems \cite{h6,A1} and in the design of coupled laser systems \cite{A2,A3,A4,A5,A6}. In a landmark work \cite{U1}, Gong and collaborators reconsidered the Hatano-Nelson model. They unravelled that in the clean (disorder-free) case the model displays a nontrivial point-gap topology, inherent to the energy spectrum in 
 complex plane, whereas in the disordered system the non-Hermitian delocalization transition induced by the imaginary gauge field has a topological origin. NH delocalization transitions associated to spectral and topological phase transitions have been 
  predicted in many other NH models as well,
such as in  NH extensions of the Aubry-Andr\'e model  \cite{h4,h7,h8,h9,h14,h15,h16,h19,h20,h27,h32,h35,h36,h37,h54} displaying aperiodic order rather than uncorrelated disorder, culminating with their recent experimental demonstrations in synthetic photonic lattices \cite{e5,e6}.\\

In this work we investigate the spectral and localization/delocalization phase transitions induced by an imaginary gauge field in NH superlattices, where the unit cell of the superlattice contains a large number $M$ of sites. In a superlattice, strictly speaking 
all eigenstates are extended under periodic boundary conditions (PBC), and a spectral phase transition, from an entirely real energy spectrum to complex energies, does not strictly exist. However, the motivation to 
investigate the spectral and localization features in superlattices subjected to an imaginary gauge field is twofold. On the one hand, 
a superlattice with large $M$ can serve as an approximation of a lattice with aperiodic order \cite{Th1,Th2}, and thus could provide a framework to unravel how the spectral and NH delocalization phase transitions arise in disordered lattices.
 This entails us to introduce the concept of imperfect (or smooth) phase transitions, as it is observed in statistical physics in systems with a finite number of degrees of freedom (or particles), where first order phase transitions arise when going from finite-size closed systems to the thermodynamic limit \cite{S1}: while we can never generate a sharp transition with a finite number $M$ of sites per unit cell, as we increase $M$ the spectral and localization features of the system become sharper and sharper, and a truly discontinuous transition is observed in the infinite $M$ limit (infinite degrees of freedom). 
 On the other hand, the superlattice system allows us to disclose that  in the absence of disorder a spectral phase transition can still be observed in the large $M$ limit, which however {\em does not} correspond to a non-Hermitian delocalization phase transition. 
 In other words, the coincidence of spectral and localization/delocalization phase transitions previously predicted and observed in NH models strictly requires some kind of disorder in the system.
 The predicted phenomena are illustrated by considering specific superlattice models and could be observed in non-Hermitian photonic quantum walks, where synthetic superlattices with controllable $M$ and imaginary gauge fields can be realized with existing experimental apparatus.

%Phase transitions are a many-body effects. You can not generate sharp transition with a finite number of degrees of freedom (or particles). However as you add particles the features of the system may become sharper. In the limit of infinitely many particles (thermodynamic limit) you get a truly discontinuous transition.

\section{Spectral and localization properties of non-Hermitian superlattices with an imaginary gauge field} 
\subsection{Model}
The starting point of our analysis is provided by the Hatano-Nelson model \cite{h1,h2,U1}, describing the hopping dynamics on a one-dimensional tight-binding lattice with a superimposed imaginary magnetic flux.
In physical space, the dynamics is described by the  Schr\"odinger equation 
\begin{equation}
i \frac{d \psi_n}{dt}= J \exp(h) \psi_{n+1}+J \exp(-h) \psi_{n-1}+V_n \psi_n
\end{equation}
for the wave function amplitude $\psi_n$ at the $n$-th site in the lattice, where $J$ is the hopping amplitude, $h$ is the imaginary gauge field, and $V_n$ is the real on-site potential.  For $h \neq 0$, the Hamiltonian of the system is NH as the imaginary 
gauge field introduces an asymmetry between left $J_L=J \exp(h)$ and right $J_R=J \exp(-h)$ hopping amplitudes. In the clean limit $V_n=0$, the system displays the NH skin effect \cite{U4,U5,U6,U7,U8,U9,U10,U11,U12,U13}, i.e. the strong dependence of the energy spectrum and corresponding eigenstates on the boundary conditions. In the following we will always assume  an arbitrarily large system size and PBC, which are necessary to observe the phenomenon of NH delocalization transition in 
disordered systems \cite{h1,h2,U1}.\\
Let us first briefly recall the phenomenon of NH delocalization \cite{h1,h2}. Let us assume that $V_n$ describes a disordered potential, with $V_n$ either a random uncorrelated potential (like in the Anderson model) or an incommensurate potential (like in the Aubry-Andr\'e model). For $h=0$ and for a sufficiently strong disorder, all eigenstates of the Hamiltonian with real energy $E$ are exponentially localized with a localization length $1/ \gamma(E)$ (inverse of the Lyapunov exponent). Let us indicate by $\gamma_{m}>0$ the minimum value of the Lyapunov exponent $\gamma(E)$ among all eigenstates of the Hamiltonian at $h=0$, i.e. $\gamma_{m}= {\rm min}_{E} \gamma(E)$. As $h$ is increased above zero, for $h<\gamma_m$ the energy spectrum remains entirely real, it is not modified by the imaginary gauge field, and all eigenstates remain localized: the main effect of the gauge field is to introduce an asymmetry in the localization between left and right tails of the wave functions. On the other hand, for $h> \gamma_m$ the energy spectrum changes and some (or even all) energies become complex. Interestingly, the eigenstates corresponding to complex energies become delocalized: this is basically the phenomenon of NH delocalization predicted by Hatano and Nelson. For example, for the incommensurate potential $V_n=2 V \cos(2 \pi \alpha n)$, with $\alpha$ irrational Diophantine and $V>J$, the NH delocalization transition is observed at $h=\gamma_m=\log (V/J)$, with all wave functions undergoing a simultaneous delocalization transition for $h>\gamma_m$  (see for instance \cite{h13}).\\
 \begin{figure}[htbp]
  \includegraphics[width=86mm]{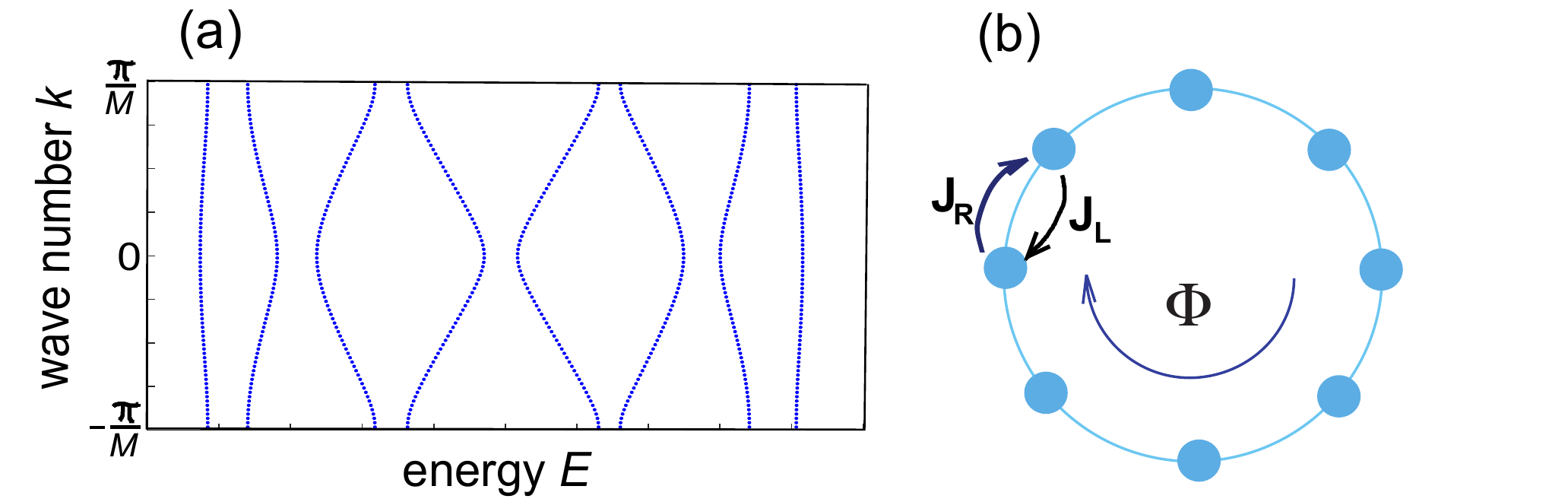}\\
  \caption{(color online) (a) Schematic of the energy spectrum of a superlattice in the Hermitian limit $h=0$ ($J_L=J_R=J$) under PBC. The spectrum comprises $M$ minibands ($M=8$ in the specific example of the figure), with dispersion relations $E=E_l(k,h)$ ($l=1,2,..,M$) defined by the eigenvalues of the Bloch Hamiltonian $\mathcal{H}(k)$ [Eq.(4)]. The Bloch wave number $k$ varies in the range $(-\pi/M, \pi/M)$. (b) The superlattice with PBC can be mapped onto a ring lattice, comprising $M$ sites, with a superimposed magnetic flux $\Phi=Mk$.}
\end{figure}
Here, we consider instead the case of a superlattice comprising a large number $M$ of sites in each unit cell. This corresponds to assume
\begin{equation}
V_{n+M}=V_n
\end{equation}
for the on-site real potential. 
Clearly, the case of a disordered lattice can be obtained from the superlattice model in the limiting case $M \rightarrow \infty$ \cite{Th1}.\\
 In a superlattice, the eigenfunctions with energy $E$ are delocalized Bloch states of the form $\psi_n(t)= \phi_n \exp(-iEt)$ with $\phi_n$ satisfying the periodicity condition
\begin{equation}
\phi_{n+M}=\phi_n \exp(ikM)
\end{equation}
where $-\pi/M \leq k < \pi/M$ is the Bloch wave number (quasi-momentum). The Bloch Hamiltonian $\mathcal{H}(k)$ of the superlattice is given by the $M \times M$ matrix
\begin{widetext}
\begin{equation}
\mathcal{H}(k)= \left(
\begin{array}{ccccccc}
V_1 & J_L & 0 & ...& 0 & 0& J_R \exp(-ikM) \\ 
J_R & V_2 & J_L & ... & 0 & 0& 0 \\
0 & J_R & V_3 & ... & 0 & 0 & 0 \\
... & ... & ... & ... & ... & ... & ... \\
0 & 0 & 0 & ... & J_R & V_{M-1} & J_L \\
J_L \exp(ikM) & 0 & 0 & ... & 0 & J_R & V_M
\end{array}
\right)
\end{equation}
\end{widetext}
and the corresponding energy spectrum is described by $M$ minibands with dispersion curves $E=E_l(k,h)$ ($l=1,2,3,...,M$), which are obtained from the eigenvalues of $\mathcal{H}(k)$; see Fig.1(a).
We note that the energy spectrum of the Bloch Hamiltonian can be mapped onto the energy spectrum of a ring lattice with $M$ sites, hopping amplitudes $J_{L,R}$ and on-site potential $V_n$ ($n=1,2,...,M$), crossed by a magnetic flux $\Phi=Mk$, as schematically shown in Fig.1(b).\\
The localization properties of the Bloch wave function of energy $E$ {\em within each unit cell} of the superlattice, i.e. in the corresponding ring lattice of Fig.1(b), can be captured by the inverse participation ratio (IPR)
\begin{equation}
{\rm IPR}(E)= \frac{\sum_{n=1}^{M} |\phi_n|^4}{\left( \sum_{n=1}^{M} | \phi_n|^2 \right)^2},
\end{equation}
with ${\rm IPR} \sim 1$ for a tightly localized wave function, and ${\rm IPR} \sim 1/M \ll 1$ for a delocalized wave function. For the whole set of Bloch wave functions, the localization properties of the superlattice within each unit cell can be 
summarized by three IPR values: the largest IPR value among all wave functions, indicated by IPR$_{max}$ and corresponding to the IPR of the most localized wave function;  the smallest IPR value among all wave functions, indicated by IPR$_{min}$ and corresponding to the IPR of the most extended wave function; and the mean value of the IPR among all wave functions, indicated by $\overline{{\rm IPR}}$.\\
 It can be readily shown that, since the eigenvalues of the matrix $\mathcal{H}$ do not change after a similarity transformation, the energy bands in the non-Hermitian case $h \neq 0$ can be obtained from the ones in the Hermitian case $h=0$ by a complexification of $k$, namely one has
\begin{equation}
E_l(k,h)=E_l(k-ih,0).
\end{equation}
\subsection{Spectral phase transitions}
Since $E_l(k,0)$ is given by a sum over $n$ of terms oscillating like $\sim \cos(kMn)$, it is clear that for any non-vanishing value of $h$ the energy spectrum cannot remain strictly real after the substitution $k \rightarrow k-ih$, according to Eq.(6). Therefore, strictly speaking in a superlattice with finite $M$ there is not any spectral phase transition, from real to complex energies, as $h$ is increased above zero. However, akin to what happens in statistical physics where sharp phase transitions only emerge when going from finite-size closed systems with a finite number of particles to the thermodynamic limit \cite{S1}, a main question is whether a sharp spectral phase transition, from an entire "quasi" real energy spectrum to complex one, can nevertheless emerge in a superlattice in the large $M$ limit. This possibility is clearly expected when the on-site potential sequence $V_n$ approximate a disordered lattice
with a sufficient strong degree of disorder \cite
{Th1}. It is clear that for an arbitrary choice of the sequence $V_n$ this is not the case, i.e. the imaginary part of the eigenenergies smoothly deviates from zero as $h$ is increased, and no sharp transition emerges in the large $M$ limit.  This is shown, as an example, in Fig.2, which depicts the numerically-computed behavior versus $h$ of the largest part of $| {\rm Im} E_l(k,h)|$, over both $k$ and $l$, for the on-site potential sequence defined by the sequence
\begin{equation}
V_n=
\left\{
\begin{array}{cl}
A & n=1 \\
0 & n=2,3,...,M
\end{array}
\right.
\end{equation} 

 \begin{figure}[htbp]
  \includegraphics[width=86mm]{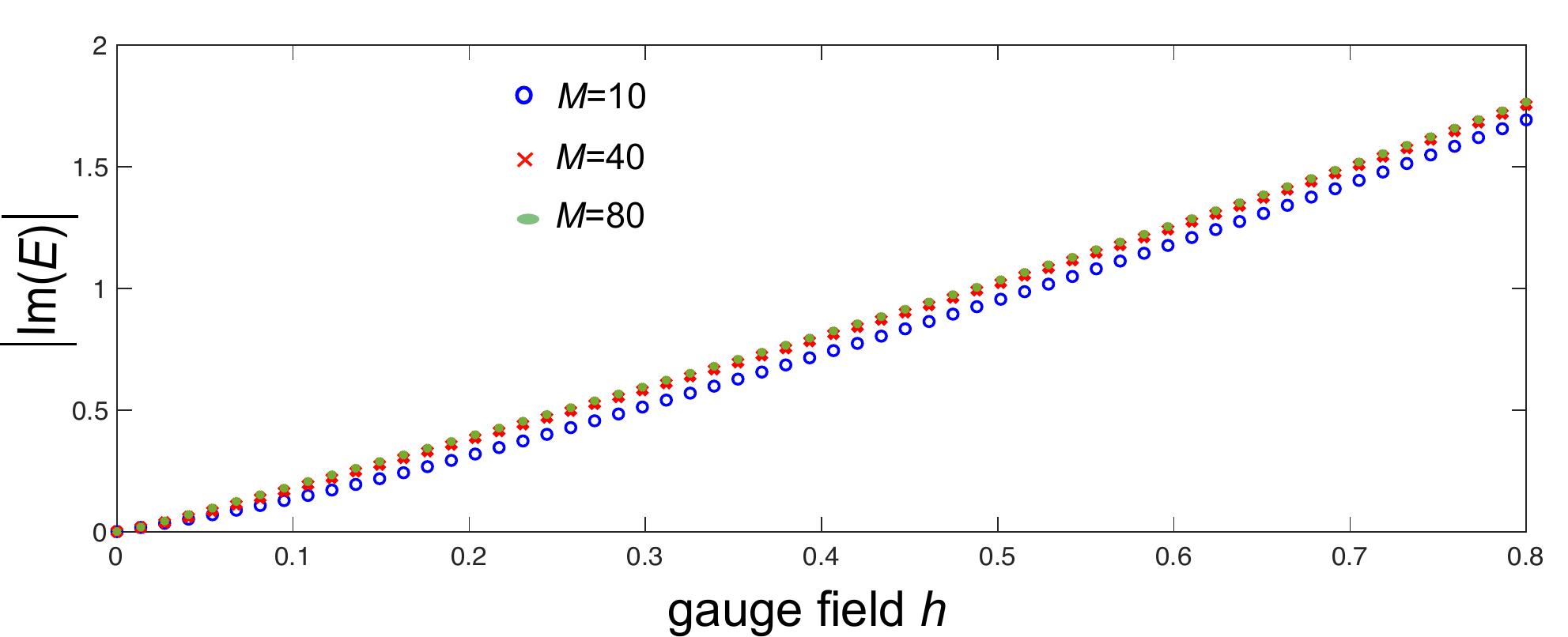}\\
  \caption{(color online) Numerically-computed behavior of the largest value of the modulus of the imaginary part of energy, i.e. ${\rm max}_{k,l} |{\rm Im}(E_l(k,h))|$, versus $h$ for the superlattice defined by the sequence (7) with $A=2.5$ and for a few increasing values of $M$ ($M=10,40$ and 80).  The hopping rate is $J=1$. Note that the imaginary part of energy smoothly increases with $h$ and no sharp phase transitions are observed as $M$ is increased.}
\end{figure}
\begin{figure}[htbp]
  \includegraphics[width=86mm]{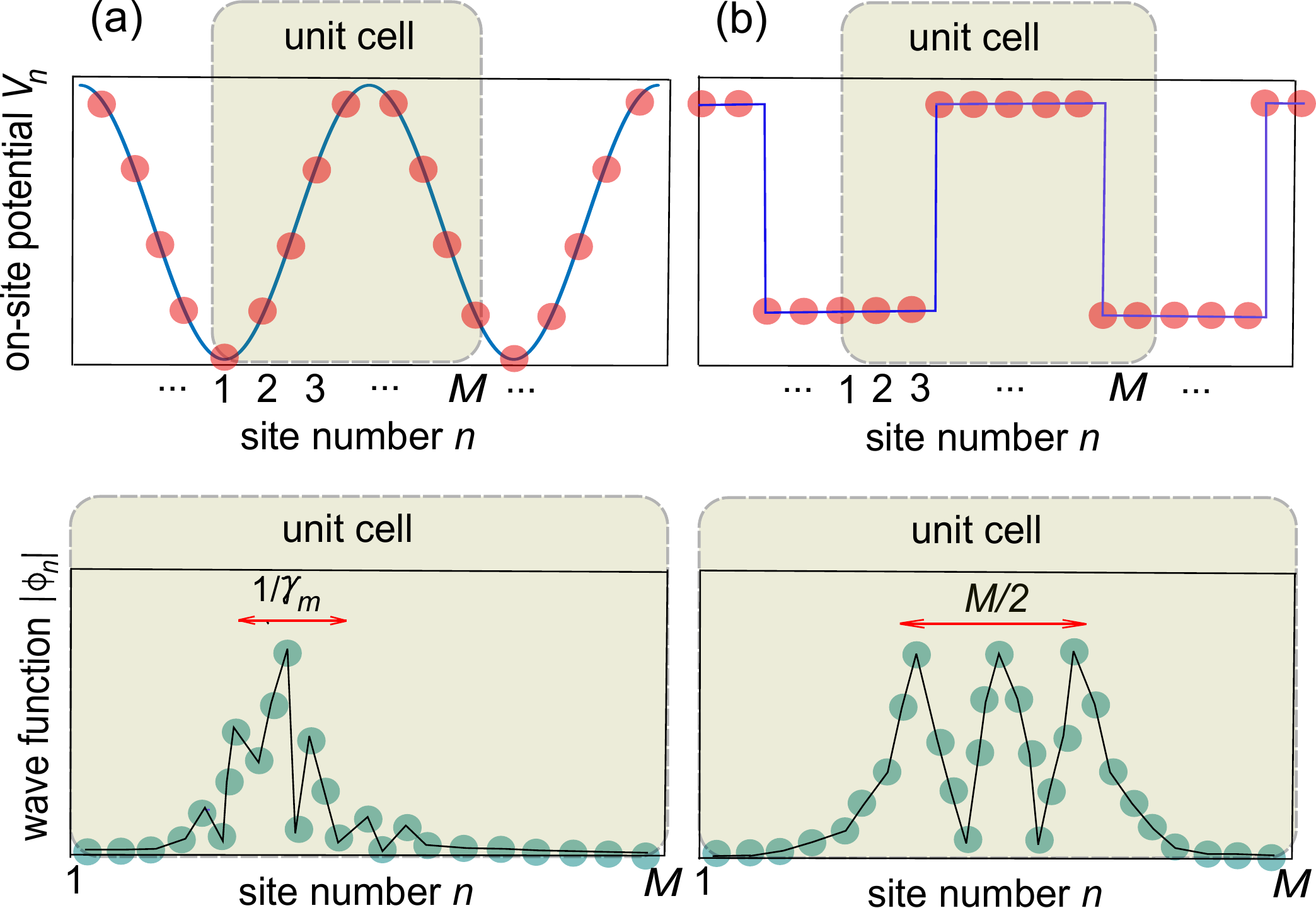}\\
  \caption{(color online) Examples of superlattices displaying flat bands in the infinite $M$ limit. (a) Superlattice with on-site periodic potential $V_n$ defined by the sequence Eq.(9). In the large $M$ limit, the superlattice realizes the Aubry-Andr\'e incommensurate potential. For $V>J$ and $h=0$,  all wave functions are tightly localized within each  unit cells (bottom panel) with a localization length $1 / \gamma_m$ which is independent of $M$ and given by $1 / \log (V/J)$. (b) Superlattice with on-site potential defined by the sequence Eq.(11). The potential realizes a sequence of potential barriers, each of width $M/2$. For $h=0$, the wave functions in each unit cell are tightly localized (evanescent waves with exponential decay) in one half of the unit cell, and extended (oscillating) in the other half of the unit cell (bottom panel). While in (a) the IPR of the wave function in each unit cell [Eq.(5)] remains finite as $M \rightarrow \infty$, in (b) the IPR vanishes as $M \rightarrow \infty$.}
\end{figure}
Clearly, as $M$ increases, no sharp phase transition emerges for this sequence.
However, let us suppose that at $h=0$ the sequence $V_n$ is chosen such that the $M$ minibands of the superlattice become flatter and flatter as $M$ is increased. This scenario usually happens when each of the $M$ Bloch wave functions $\phi_n^{(l)}$ in the $l-th$ miniband displays exponentially-decaying tails in each unit cell with a characteristic decay rate $\gamma_m$, as schematically depicted in Fig.3. In this case, the dispersion curves $E_l(k,0)$ of the nearly-flat bands can be calculated by a perturbative analysis and read 
 $E_l(k,0) \simeq E_l(0,0)+\Delta_l [1-\cos(kM)]$, where the bandwidth $2 \Delta_l$ is  given in terms of the product of the small-amplitude exponentially-decaying wave function at the edge sites of the unit cell; technical details are given in Appendix A. 
Specifically, for large $M$ the bandwidth $\Delta_l$ vanishes with $M$ as 
 \[ \Delta_l \sim  \exp(-\sigma_l M), \]
 where $\sigma_l= \rho_l \gamma_m$ and $\rho_l$, with $ 0 < \rho_l \leq 1$, is the fraction of the unit cell where the wave function $\phi_n^{(l)}$ is evanescent (see Appendix A for technical details). For example, for the potential of Fig.3(a) one has $\rho_l=1$, whereas for the potential of Fig.3(b) one has $\rho_l=1/2$.\\ 
 When we apply a non-vanishing imaginary gauge field $h>0$ in such a nearly-flat band system, 
 the dispersion curve of the $l-th$ band 
 is simply obtained by the replacement $k \rightarrow k-ih$; in the large $M$ limit one then obtains the following scaling law
 \begin{eqnarray}
 E_l(k,h)-E_l(0,0)  & \sim \frac{1}{2}  \exp[ikM+(h-\sigma_l)M]
\end{eqnarray}
i.e. $E_l(k,h)$  is approximately described by a circle in complex energy plane, centered at the real energy $E_l(0,0)$ and of radius that scales with $M$ as $ \sim (1 /2) \exp[(h-\sigma_l)M]$.
Therefore, the largest imaginary part of the energy of the $l-th$ miniband displays a sharp change, from nearly zero to large values, as $h$ is increased above the critical value $\sigma_l$: in fact, in the large $M$ limit one has
 $\exp[(h-\sigma_l)M] \ll 1$ for $h<\sigma_l$ and  $\exp[(h-\sigma_l)M]  \gg 1$ for $h>\sigma_l$.
   To sum up, we expect an imperfect spectral phase transition, which becomes sharper and sharper (i.e. exact) in the infinite $M$ limit, when $h$ reaches a critical value, defined by the smallest value of $\sigma_l$ among the various minibands.
   
    We have checked the emergence of sharp phase transitions by considering in details two specific models, schematically depicted in Fig.3. \\ 
   The first model [Fig.3(a)] realizes a commensurate approximation of the NH Aubry-Andr\'e model and is described by the sequence
   \begin{equation}
   V_n=2 V \cos ( 2 \pi \alpha_M n)
   \end{equation}
   where $\alpha_M=R/M$ is a  rational that approximates a Diophantine irrational number $\alpha$. For example, assuming $\alpha=(\sqrt{5}-1)/2=0.618033...$ (the inverse of the golden ratio),
the sequence of rationals $\alpha_M=R/M=p_{s-1}/p_s$ converges to $\alpha$ in the $s \rightarrow \infty$ limit, where $
p_s = 0,1,2,3,,5,8,13,21,34,55,89,144,.. $ are the Fibonacci numbers. Therefore, after letting $M=p_s$ and $R=p_{s-1}$, for large $s$ the superlattice with the potential given by Eq.(9) provides an approximation of incommensurate disorder.
For $V>J$, as we increase $h$ we expect an imperfect spectral phase transition, from nearly real energy spectrum to complex one. The critical value of $h$ at which the phase transition occurs can be calculated analytically \cite{h7}; in terms of the nearly-flat band analysis discussed above, we can assume $\rho_l=1$ and identify the parameter $\sigma_l$, which defines the bandwidth scaling $ \Delta_l \sim  \exp(-\sigma_l M)$ of each miniband, as the Lyapunov exponent (inverse of the localization length) $\gamma_m$ of the wave functions. For the Aubry-Andr\'e model one has $\gamma_m=\log(V /J)$, and therefore the critical value $h$ of the spectral phase transition reads 
\begin{equation}
h=\gamma_m=\log(V /J). 
\end{equation}
This behavior is illustrated in Fig.4(a), which shows the numerically-computed shape of the largest value of the modulus of the imaginary part of energy $E$ versus $h$ for a few increasing values of $M$. Note that, as expected, the phase transition becomes sharper as $M$ is increased, i.e. as the rational $\alpha_M$ gets closer to the irrational $\alpha$. {\color{black} The full energy spectrum in complex energy plane, for a few increasing values of $h$ and for $M=144$, is shown in Fig.4(c). When $h$ is increased above zero but it stays below the critical value $h_c$, the energy spectrum remains almost real and it is not affected by the imaginary gauge field: the energy spectrum is basically frozen. When $h$ is increased above the critical value $h_c$, the energies are forced to move into the complex plane and distribute along closed loops. As $h$ is further increased, the closed loops merge forming larger loops. This spectral deformation scenario is typical of non-Hermitian lattices with disorder and imaginary gauge fields, and has been discussed in several previous works (see e.g. \cite{h3g,h8,h9,h12,h53,h55}). A special feature of the sinusoidal incommensurate potential is that all energies simultaneously become complex, which is related to the well-known property that the Lyapunov exponent of eigenstates in the Aubry-Andre model does not depend on energy.}

\begin{figure}[htbp]
  \includegraphics[width=86mm]{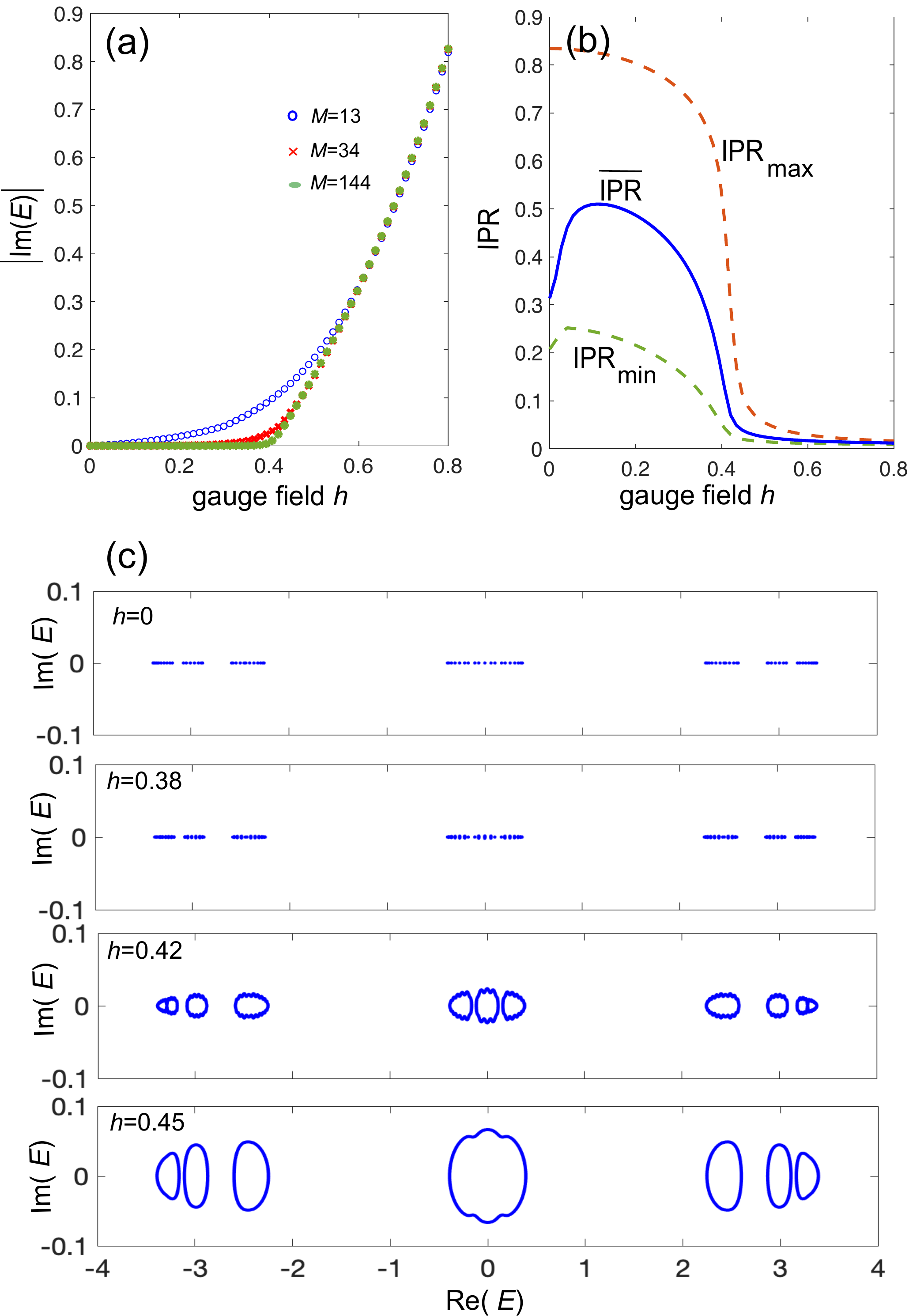}\\
  \caption{(color online) (a) Numerically-computed behavior of the largest value of the modulus of the imaginary part of energy, i.e. ${\rm max}_{k,l} |{\rm Im}(E_l(k,h))|$, versus $h$ for the superlattice defined by the sequence (9) 
  with $V=1.5$ and for a few increasing values of $M$ ($M=13,34$ and 144, corresponding to $\alpha_M=8/13, 21/34$ and 89/144, respectively).  The hopping rate is $J=1$. Note that the imaginary part of energy undergoes a sharp phase transition at the critical value $h=\log (V/J) \simeq 0.406$ in the large $M$ limit. (b) Numerically-computed behavior of the three IPR parameters (IPR$_{max}$, IPR$_{min}$ and $\overline{\rm{IPR}}$) for the superlattice with $M=144$. Note a clear phase transition from all localized states for $h< \sim 0.4$ to all extended states for $h > \sim 0.4$. {\color{black}(c) Detailed behavior of the energy spectrum in complex plane for $M=144$ and for a few increasing values of $h$. Note the appearance of closed loops in complex plane, emanating from the minibands of the superlattice on the real energy axis, as $h$ is increased above the critical value $h_c \simeq 0.406$. The loops enlarge and merge as $h$ is increased further above $h_c$.}}
\end{figure}
   
\begin{figure}[htbp]
  \includegraphics[width=86mm]{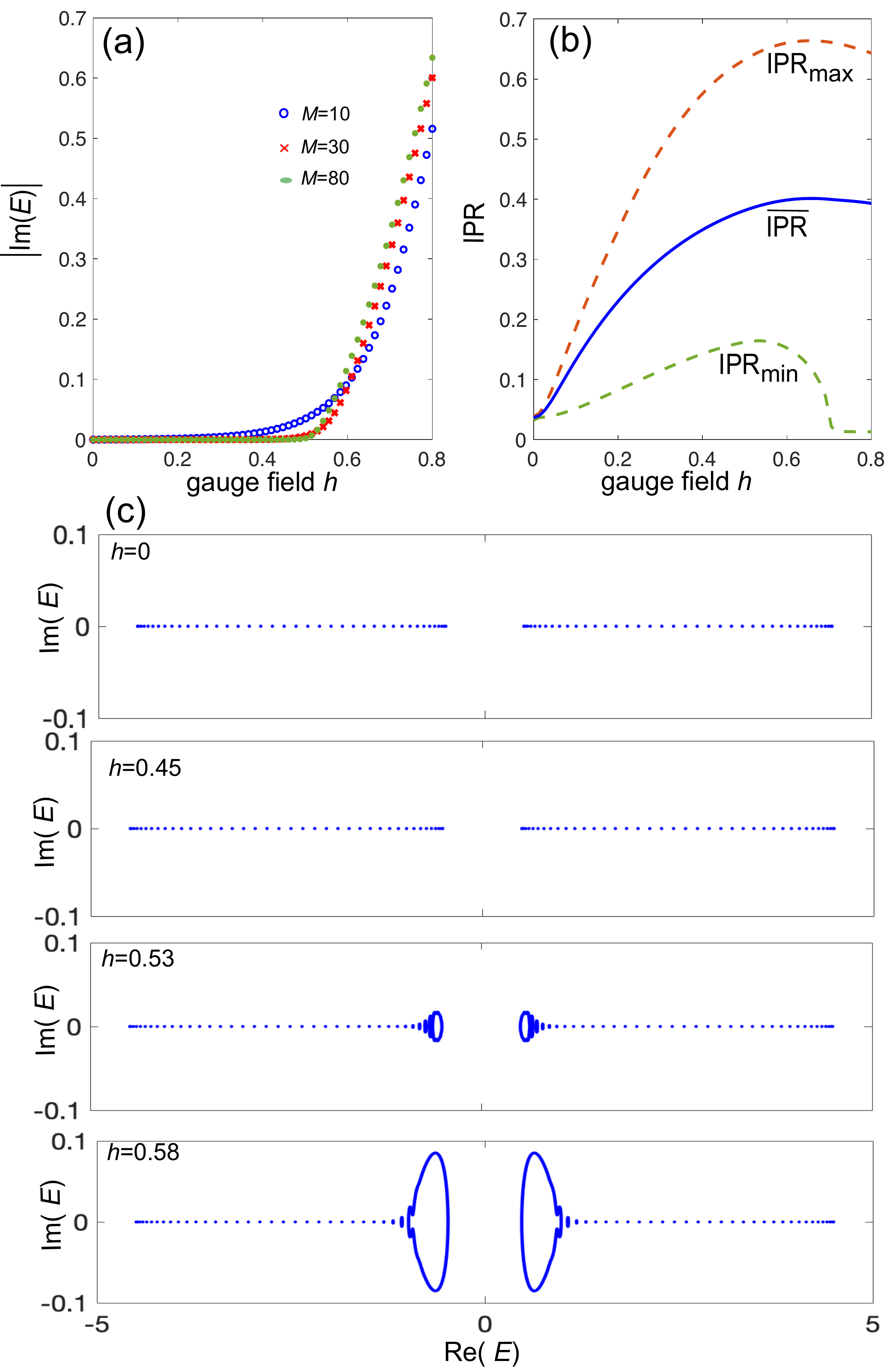}\\
  \caption{(color online) Same as Fig.4, but for the ordered superlattice defined by the potential sequence Eq.(11). Parameter values are $J=1$ and $V=2.5$. A clear phase transition in the energy spectrum [panel (a)] is observed in the large $M$ limit at the critical value $h \simeq (1/2) {\rm acosh} (V/J-1) \simeq 0.48$ of the imaginary gauge field. However, no special features are observed in the localization properties of the wave functions near the critical value of $h$ [panel (b)]. {\color{black} The spectral curves in (c) are numerically computed assuming $M=80$. The deformation features of the energy spectrum, as the phase transition point $h=h_c$ is crossed, are similar to the ones shown in Fig.4(c).}}
\end{figure}

   The second model [Fig.3(b)] is an ordered superlattice consisting of a sequence of rectangular potential barriers on the lattice and described by the sequence
   \begin{equation}
   V_n=
   \left\{
   \begin{array}{cl}
   V & n=1,2,..., M/2 \\
   -V & n=M/2+1,M/2+2,...,M 
   \end{array}
   \right.
   \end{equation}
where we assumed an even value of $M$ and $V>2J$. Note that such a superlattice can be also viewed as a sequence of interfaced crystals (heterojunction), each of size $M/2$ (in lattice period units) and with non-overlapped tight-binding bands $2 J \cos k \pm V$, i.e shifted one another by $ 2V$. For $h=0$, $V \gg 2J$ and in the infinite $M$ limit, such a superlattice displays two branches of $M/2$ flat bands with energies $E_l^{\pm}=\pm V+ 2J \cos [2 \pi l/(M+2)]$ ($l=1,2,...,M/2)$. The two branches of flat bands, with either the plus or minus sign, correspond to Bloch wave functions which are extended (oscillating) in one half of the unit cell, and exponentially-damped (evanescent) in the other half of the unit cell, as schematically shown in the bottom panel of Fig.3(b).  As $h$ is increased, a sharp phase transition is observed in the energy spectrum in the large $M$ limit, as shown in Fig.5(a).  {\color{black} The full energy spectrum in complex energy plane, for a few increasing values of $h$ and for $M=80$, is shown in Fig.5(c). Like for the previous model, when $h$ is increased above zero, but it stays below the critical value $h_c$, the energy spectrum remains almost real and it is not affected by the imaginary gauge field. When $h$ is increased above the critical value $h_c$, some energies are forced to move into the complex plane and distribute along closed loops, while others remain almost on the real axis. As $h$ is further increased, a larger number of energies becomes complex forming closed loops, and adjacent loops enlarge and merge forming wider loops.}  Similar to the Aubry-Andr\'e model discussed above, the critical value of $h$, above which the imaginary part of the energy spectrum sharply starts to increase, can be computed by letting $h=\sigma_l$, where $\sigma_l$ is the exponent entering in the bandwidth $2 \Delta_l$ of the miniband. Since the wave function in Fig.3(b) is evanescent solely in a half of the unit cell, we can assume $\rho_l=1/2$ and thus $\sigma_l= \gamma_m/2$, where  $\gamma_m$ is the decay rate of the evanescent (exponential) tails of the wave function. The value of $\gamma_m$ can be readily calculated from a simple eigenvalue analysis of a rectangular potential barrier (a junction) on a tight-binding lattice, yielding
\begin{equation}
h=\frac{\gamma_m}{2}= \frac{1}{2}{\rm acosh} \left( \frac{V}{J}-1 \right).
 \end{equation}
   \subsection{Spectral phase transition and localization}

   The natural question is whether the spectral phase transition discussed in the previous subsection, arising from the band flattening in the large $M$ limit, is also associated with a change in the localization properties of the wave functions, as one would expect for a disordered lattice. 
  The answer to this question is that, while for a disordered or incommensurate sequence $V_n$ the emergence of complex energies corresponds to the usual NH delocalization transition like in the original Hatano-Nelson model, in the most general case the localization properties of the wave functions do not undergo any significant change when crossing the spectral phase transition point. In other words, the coincidence of spectral and NH delocalization transitions requires some kind of disorder in the system. To illustrate such a main result, let us consider the two models of Figs.3(a) and (b), which display both a spectral phase transition in the large $M$ limit when the gauge field $h$ is increased above a critical value. In the incommensurate potential of Fig.3(a), the spectral phase transition is associated to the usual NH delocalization transition: the wave functions associated to complex energies become extended. This is shown in Fig.4(b), which depicts the numerically-computed behavior of the three IPR parameters (the minimum, maximum and average values of IPR of the wave functions within one unit cell) versus $h$. As one can clearly see, in proximity of the spectral phase transition, i.e. around the critical value $h$ given by Eq.(10), the IPR parameters undergo a rapid change from finite values (in the localized phase) to small values (in the extended phase). However, when considering the superlattice model of Fig.3(b), corresponding to an ordered superlattice with alternating potential barriers, as $h$ is varied across the critical value, given by Eq.(12), the IPR parameters do not show any special change, as shown in Fig.5(b), i.e. there is not any NH localization/delocalization transition here associated to the spectral phase transition. To qualitatively understand the behavior of the IPR versus $h$ shown in Fig.5(b), let us first consider the Hermitian limit: for $h=0$, the  eigenfunctions $\phi_n$ in the two miniband branches are extended waves in one half of the unit cell, whereas they are evanescent waves with exponential decay tails in the other half space of the unit cell, as illustrated in the bottom panel of Fig.3(b). Hence, for $h=0$ the IPR of the wave functions take small values,  scaling with $M$ like $ \sim 1 /M$. As we slightly increase the imaginary gauge field $h$, the energy spectrum does not sensitively change and the wave functions at $h \neq 0$, $\phi^{\prime}_n$, are basically obtained from those at $h=0$ by the transformation $\phi^{\prime}_n=\phi_n \exp(-hn)$. Owing to such a transformation, the oscillating (extended) nature of $\phi_n$ in half of the unit cell acquires an exponential envelope and tends to be squeezed toward the potential barrier region, which is reminiscent of the NH skin effect in systems with open boundary conditions \cite{U4}. Therefore, as $h$ is increased above zero the IPR increases, as a result of the partial localization effect introduced by the imaginary gauge field. When $h$ crosses the critical value corresponding to the spectral phase transition point ($h \simeq 0.48$ in the example of Fig.5), the IPR undergo smooth changes, and after reaching a maximum they decrease. Note that IPR$_{min}$, corresponding to the IPR of the most delocalized wave function, reaches a small value, $ \sim 1 /M$, only at $h \simeq 0.7$, i.e. well above the spectral phase transition point. At much larger values of $h$, not shown in Fig.5(b),  also IPR$_{\max}$ finally decays to a  small value, of the order $\sim 1 /M$, indicating that all eigenstates become delocalized. Therefore, in the ordered superlattice of Fig.3(b) the spectral phase transition, from a real to a complex energy spectrum, does not coincide with a delocalization of the wave functions.\\
  The seemingly counterintuitive result that a spectral phase transition can occur without a corresponding change of the localization properties of the wave functions, as originally predicted by Hatano and Nelson, can be explained by the main circumstance that, as shown by Eq.(8), a spectral phase transition only requires that at $h=0$ the system displays flat bands in the large $M$ limit, with bandwidths exponentially vanishing with the system size $M$.  However, band flattening does not necessarily require localization of the wave functions. Indeed, the  main distinctive feature between a disordered superlattice [as the one shown in Fig.3(a) and approximating an incommensurate disorder in the large $M$ limit] and an ordered superlattice [as the one shown in Fig.3(b)], both displaying band flattening in the large $M$ limit and an associated spectral phase transition, is that in a disordered system the wave functions $\phi_n$ at $h=0$ and in the large $M$ limit are exponentially localized and scale-independent (i.e. the form of the wave function becomes independent of $M$ in the large $M$ limit),  while in the ordered lattice they are scale-dependent and can remain extended over a portion of the unit cell.
  
  \section{Phase transitions in non-Hermitian photonic quantum walk superlattices}
  The phase transitions presented in the previous section in non-Hermitian superlattices can be experimentally realized in discrete-time photonic quantum walks, which have provided in the past recent years a fascinating laboratory tool for the observation of non-Hermitian phenomena in controllable synthetic matter (see for example \cite{e1a,e1d,e2,e3,e5,e6,fiber1,fiber2,fiber3} and references therein). Specifically, let us consider discrete-time quantum walks of optical pulses in coupled fiber loops that realize a synthetic mesh lattice \cite{e1d,e5,fiber1,fiber2,fiber3}. The system consists of two fiber loops of slightly different lengths  $L \pm \Delta L$ (short and long paths) that are connected by a fiber coupler with a coupling angle $\beta$, with $0< \beta < \pi/2$. Balanced optical gain and loss are applied in the short and long fiber loops, respectively. A phase modulator is inserted in one of the two loops, which  controls the pulse phase difference as they recombine in the coupler. The traveling time of light in the two loops are $T \pm \Delta T$, where $T=L/c$,  $c$ is the group velocity of light in the fiber at the probing wavelength, and $\Delta T= \Delta L/c \ll T$ is the time mismatch arising from fiber length unbalance. After each round trip, the field amplitudes $u(t)$ and $v(t)$ of the light waves in the short and long loops at a given reference plane couple each other via the fiber coupler according to time-delayed equations (see for instance \cite{fiber4}).
Considering light dynamics at discretized times $t=t_n^m=n \Delta T+m T$, where $n=0, \pm 1, \pm2 ,...$ is the site number of the synthetic lattice at various time slots and $m$ is the round-trip number, assumed to match the traveling time $T$ along the mean path length $L$, the optical field amplitudes $u_n^{(m)}$, $v_n^{(m)}$ at the discretized times $t_n^m$ in the two loops satisfy the discrete-time coupled equations 
\begin{subequations}
 \begin{eqnarray}
 u^{(m+1)}_n & = & \left[   \cos \beta u^{(m)}_{n+1}+i \sin \beta v^{(m)}_{n+1}  \right]  \exp (h+i \varphi_n ) \label{CME1} \\
 v^{(m+1)}_n & = & \left[   \cos \beta v^{(m)}_{n-1}+i \sin \beta u^{(m)}_{n-1}  \right] \exp (-h+i \varphi_n) \;\;\;\;\;\;\; \label{CME2}
 \end{eqnarray}
 \end{subequations}
 where $h$ is the gain/loss parameter and $\varphi_n$ is an $n$-dependent phase term controlled by the phase modulator (see for instance \cite{e1d}). The phase term $\varphi_n$ plays a similar role than the potential $V_n$ in the continuous-time dynamics of NH superlattices presented in the previous section. This can be formally shown by considering the continuous-time limit of the discrete-time quantum walk  \cite{ Longhi}. In the most general case, let us note that, for spatial translational invariance $\varphi_n= \varphi$ constant and under PBC, the discrete-time quantum walk defined by Eqs.(13a) and (13b) sustains two quasi energy bands with dispersion relations given by  
 \begin{equation}
 E_{\pm}(k,h)= \pm {\rm acos} \left( \cos \beta \cos (k-ih) \right)-\varphi
 \end{equation}
 where $k$ is the Bloch wave number. Note that the quasi energies are defined mod. $ 2 \pi$. Equation (14) clearly shows that a non-vanishing phase $\varphi$ just introduces a constant shift of quasi energies. Note also that the two quasi energy bands have a width $| \pi -2 \beta|$, and for $\varphi=0$ they are centered at around $ \pm \pi/2$. 
  An inhomogeneous distribution of the phases $\varphi_n$ can be thus viewed as an inhomogeneous potential, that locally shifts the position of the two quasi energies. In particular, in the quantum walk setting a superlattice with $M$ sites in each unit cell is realized by assuming for the potential
 \begin{equation}
 \varphi_{n+M}=\varphi_n.
 \end{equation}
 In a system under PBC, the eigenfunctions  of the superlattice are extended Bloch waves of the form
 \begin{equation}
 u_n^{(m)}=U_n \exp[-iE(k,h) m]  , \; \;   v_n^{(m)}=V_n \exp[-iE(k,h) m]
 \end{equation}
 where $E(k,h)$ is the quasi energy, $k$ is the Bloch wave number which varies in the interval $(-\pi/M, \pi/M)$, and the amplitudes $U_n$, $V_n$ satisfy the boundary conditions
 \begin{equation}
 U_{n+M}=U_n \exp(ikM) \; , \;\; V_{n+M}=V_n \exp(ikM).
 \end{equation}
 The dispersion relations $E(k,h)$ of the quasi energy minibands are obtained from a determinantal equation involving the $2 M$ amplitudes $U_1,V_1,U_2,V_2,...,U_M,V_M$ (see Appendix B). Like in the continuous-time problem of Sec.II, one has  $E(k,h)=E(k-ih,0)$, i.e. the energy dispersion curve in the NH case is obtained from the dispersion curve of the Hermitian quantum walk after the substitution $k \rightarrow k-ih$.  
 Note that, owing to the binary nature of the lattice, we have $2M$ quasi energy minibands, which reduce to the two bands given by Eq.(14) when $M=1$. Like in the continuous-time model discussed in Sec.II.A, the localization properties of the Bloch wave function with quasi energy $E$ in each unit cell of the superlattice are 
 determined by the IPR parameter, given by
 \begin{equation}
{\rm IPR}(E)= \frac{\sum_{n=1}^{M}\left(  |U_n|^4+|V_n|^4 \right) }{\left( \sum_{n=1}^{M} \left( |U_n|^2+|V_n|^2 \right)  \right)^2}.
\end{equation}

 The  two models introduced in Sec.II.B and shown in Figs.3(a) and (b), corresponding to an incommensurate potential and an ordered potential with rectangular barriers, can be readily implemented in the discrete-time photonic quantum walk by an appropriate tuning of the phases $\varphi_n$.\\
 The first model (incommensurate potential) is realized by letting
 \begin{equation}
 \varphi_n= 2 \pi \alpha_M n
 \end{equation}
 where $\alpha_M=R/M$ is a rational approximant of an irrational $\alpha$. This model in the Hermitian limit $h=0$ was introduced in Refs.\cite{Ch1,Ch2,Ch3} and dubbed {\em electric quantum walk}, in analogy with the problem of Bloch oscillations on a lattice under a dc field.  The dispersion curves $E=E_l(k)$ of the quasi energy minibands for this model can be calculated in an exact analytical form \cite{Ch2}, and the explicit form depends on whether the number of sites per unit cell $M$ is odd or even. Assuming for example and odd value of $M$, one has \cite{Ch2}
 \begin{equation}
 E_l(k)=\frac{2 \pi l}{M} + (\cos \beta)^M \cos(kM)
 \end{equation}
 ($l=1,2,...,M$). 
In the non-Hermitian case, the dispersion curves are simply obtained from Eq.(20) by the replacement $k \rightarrow k-ih$. Therefore, in the large $M$ limit, a spectral phase transition, from real to complex quasi energies, is observed when $h$ is increased above the critical value
\begin{equation}
h=- \log | \cos \beta |.
\end{equation}
\begin{figure}[htbp]
  \includegraphics[width=86mm]{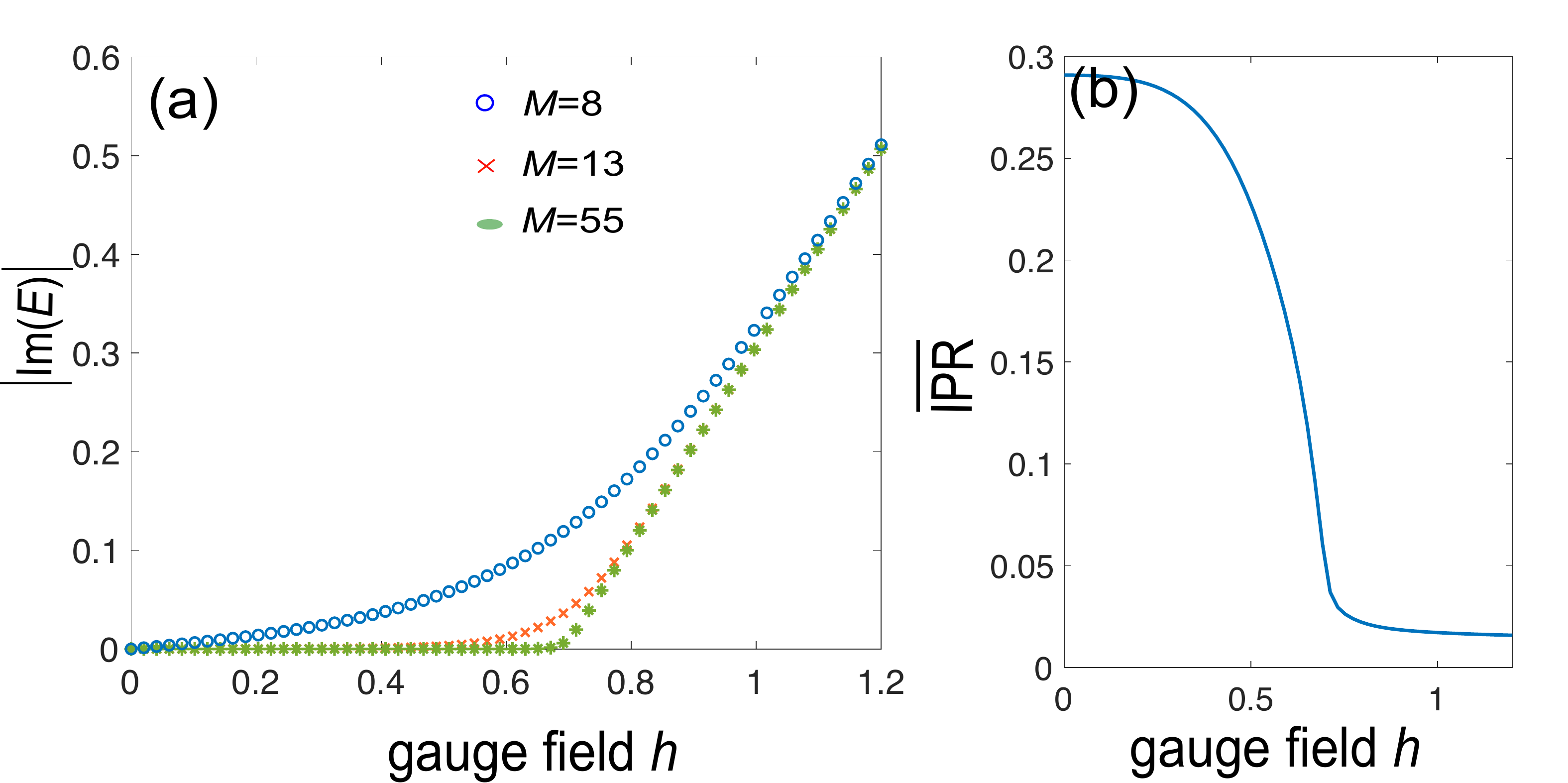}\\
  \caption{(color online) (a) Numerically-computed behavior of the largest value of the modulus of the imaginary part of the quasi energy, i.e. ${\rm max}_{k,l} |{\rm Im}(E_l(k))|$, versus $h$ for the quantum walk on a superlattice defined by the sequence (19) for $\beta= \pi/3$ and for a few increasing values of $M$ ($M=8,13$ and 55, corresponding to $\alpha_M=5/8, 8/13$ and 34/55, respectively). Note that the imaginary part of the quasi energy undergoes a sharp phase transition at the critical value $h=-\log | \cos \beta |  \simeq 0.6931$ in the large $M$ limit. (b) Numerically-computed behavior of the mean IPR parameter, $\overline{\rm{IPR}}$, for the superlattice with $M=55$. Note a clear phase transition from localized to extended states as $h$ is increased above the critical value.}
\end{figure}
 Like for the continuous-time  model, the spectral phase transition corresponds to a NH delocalization transition, from  eigenstates exponentially localized to delocalized states as $h$ is increased above the critical value (21) \cite{e5}. 
  This behavior is clearly illustrated in Fig.6, where a sequence of integers $M$ is assumed so that $\alpha_M=R/M$ is a rational approximant sequence of the inverse of the golden ratio $\alpha=( \sqrt{5}-1)/2$. Panel (a) in Fig.6 shows the numerically-computed  behavior of the largest value of the modulus of the imaginary part of quasi energy $E$ versus $h$ for a few increasing values of $M$. Note that the phase transition becomes sharper as $M$ is increased, i.e. as the rational $\alpha_M$ gets closer to the irrational $\alpha$. The spectral phase transition corresponds to a localization-delocalization transition of the wave functions in each unit cell, as clearly demonstrated in Fig.6(b) by a rather abrupt change of the mean IPR parameter as $h$ crosses the critical value given by Eq.(21).

   The second model, corresponding to an ordered superlattice and consisting of a sequence of rectangular potential barriers, is described by the phases
   \begin{equation}
   \varphi_n=
   \left\{
   \begin{array}{cl}
   V & n=1,2,..., M/2 \\
   -V & n=M/2+1,M/2+2,...,M 
   \end{array}
   \right.
   \end{equation}
 where we assumed an even value of $M$. To ensure that the Bloch wave functions of the superlattice display exponential decaying tails in half of the unit cell, we require that the quasi energy bands, as given by Eq.(14) with $\varphi= \pm V$, do not overlap. Such a condition is met whenever $\pi/2 -\beta<V<\beta$, which necessarily requires a coupling angle $ \beta > \pi/4$.
 \begin{figure}[htbp]
  \includegraphics[width=86mm]{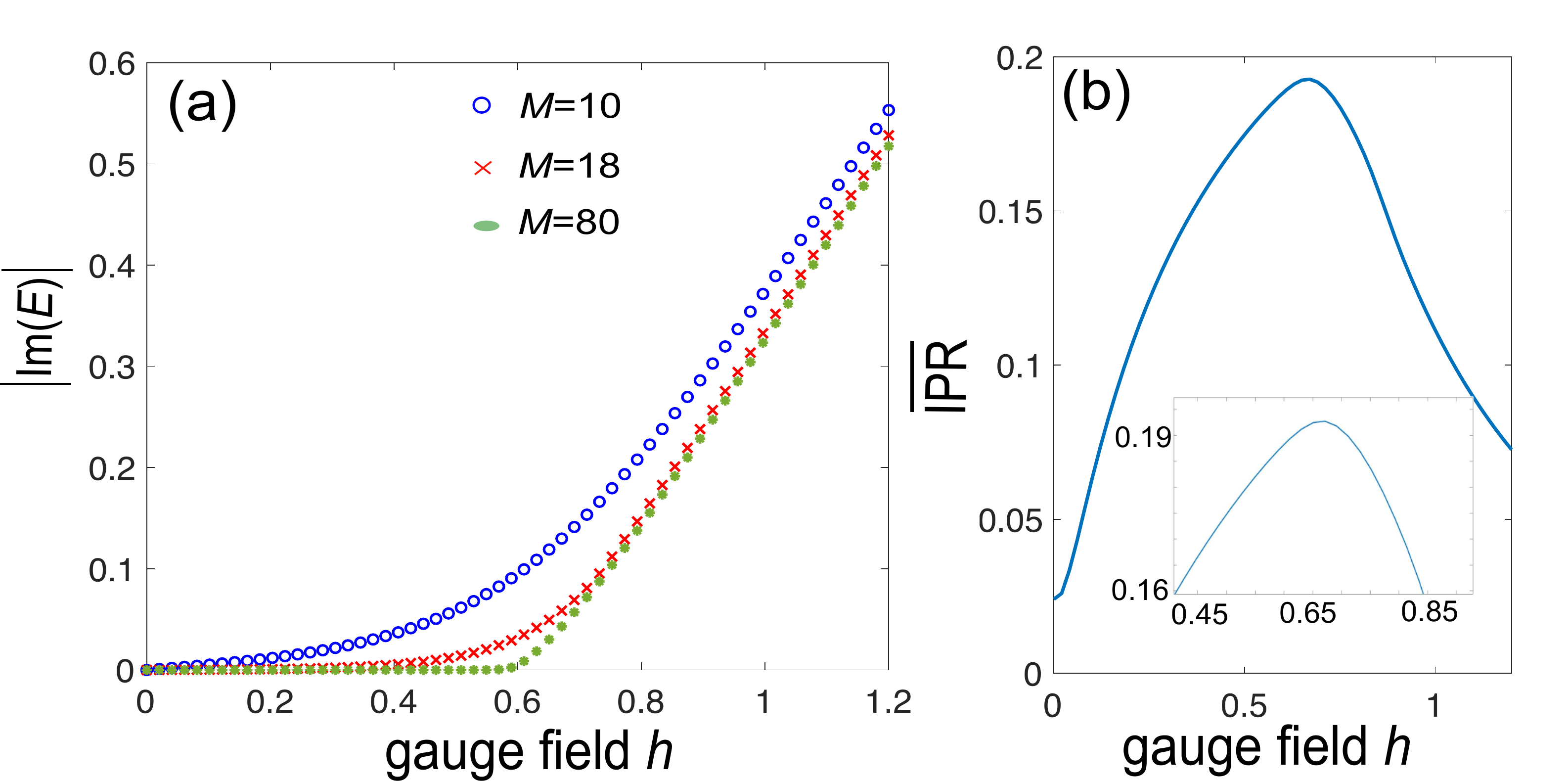}\\
  \caption{(color online) Same as Fig.6,  but for the superlattice defined by the potential sequence Eq.(22). Parameter values are $\beta= \pi/3$ and $V=\pi/4$. A clear phase transition in the energy spectrum [panel (a)] is observed in the large $M$ limit at the critical value $h \simeq 0.58$ of the imaginary gauge field, in agreement with Eq.(23). However, no special features are observed in the localization properties of the wave functions near the critical value of $h$, as indicated by the behavior of the mean value of the IPR versus $h$ [panel (b)]. The inset in (b) shows an enlargement of the mean value of the IPR curve near its maximum, which is reached at $h \simeq 0.65$, i.e. well above the critical value $h \simeq 0.58$.}
\end{figure}
Like in the second model presented in Sec.II.B, in the large $M$ limit a spectral phase transition arises as $h$ is increased above the critical value $h= \gamma_m/2$, where  $\gamma_m$ is the decay rate of the evanescent (exponential) tails of the wave function. The value of $\gamma_m$ can be readily calculated from a simple eigenvalue analysis of a rectangular potential barrier using the dispersion relation Eq.(14), yielding
\begin{equation}
h=\frac{\gamma_m}{2}= \frac{1}{2}{\rm acosh} \left( \frac{\cos(\pi-\beta-2V) }{\cos \beta} \right).
 \end{equation}
However, the spectral phase transition does not correspond to a change of the localization properties of the wave functions. This behavior is illustrated in Fig.7.\\
{\color{black}
\begin{figure}[htbp]
  \includegraphics[width=86mm]{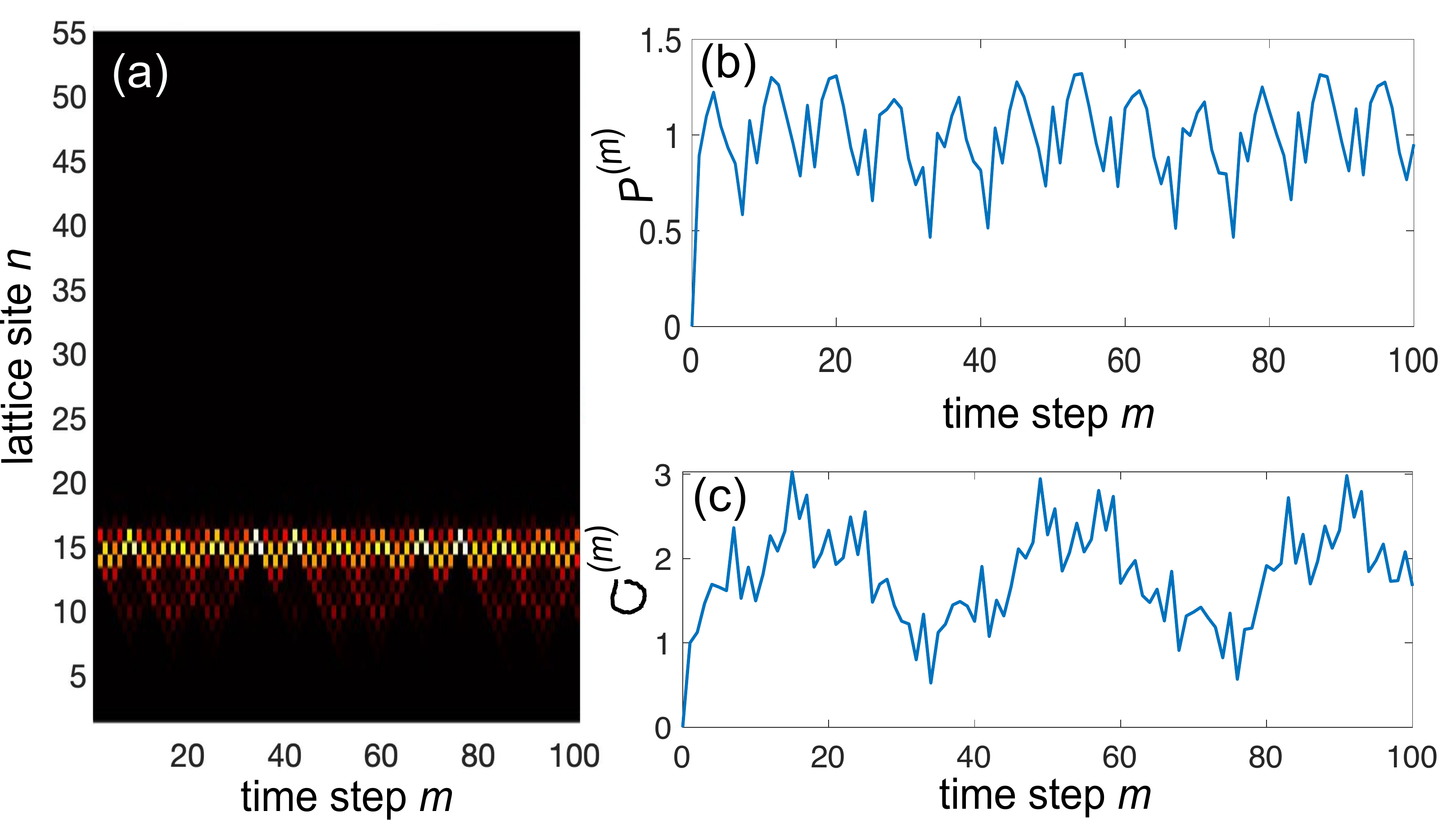}\\
  \caption{(color online)  Dynamical signatures of the phase transition in the photonic quantum walk model, defined by the potential sequence given by Eq.(19) (incommensurate potential model). Parameter values are as in Fig.6, with a number of sites $M=55$ and imaginary gauge field $h=0.4$, below the critical value $h_c=0.6931$. The system is initially excited at the site $n=n_0=14$. (a) Numerically-computed behavior of the evolution of the normalized intensity distribution $(|u_n^{(m)}|^2+|v_n^{(m)}|^2)/P^{(m)}$ on a pseudo color map in the $(n,m)$ plane. (b) Behavior of the total optical power $P^{(m)}$. (c) Behavior of the square root of the normalized second-order moment $\sigma^{(m)}= \sqrt{M_2^{(m)}}$. }
\end{figure}
\begin{figure}[htbp]
  \includegraphics[width=86mm]{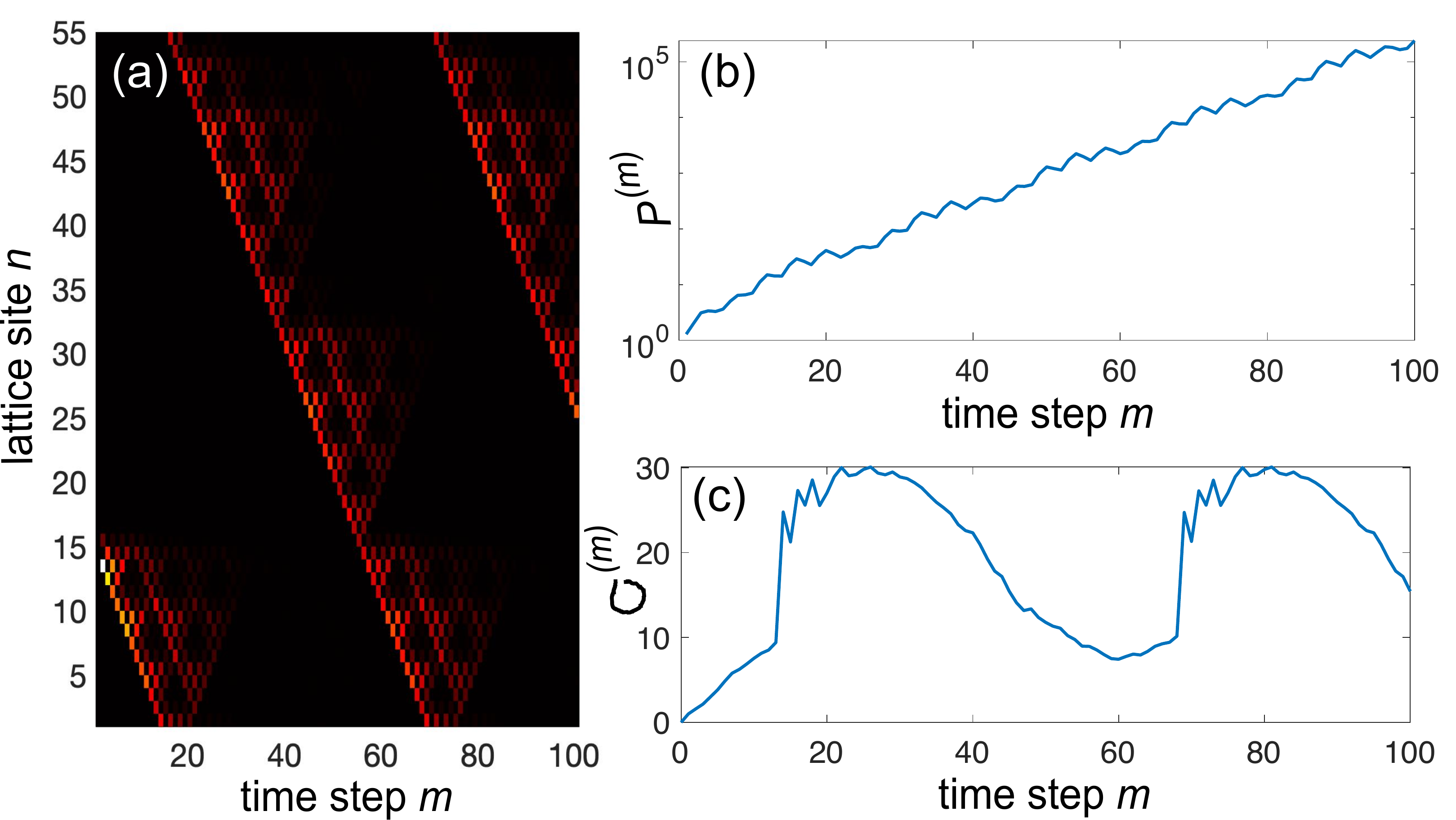}\\
  \caption{(color online) Same as Fig.8, but for $h=0.75$, above the critical value $h_c=0.6931$. The total optical power $P^{(m)}$ in (b) is plotted on a log scale.}
\end{figure}
In a photonic quantum walk experiment, the distinct spectral and localization signatures in the two types of phase transitions  can be investigated by monitoring some dynamical variables of the system, which are much more accessible than spectral quantities. Let us assume that at initial step $m=0$ we excite the superlattice in a single site (i.e. with a single optical pulse), so that $u_{n}^{(0)}=\delta_{n,n_0}$ and $v_{n}^{(0)}=0$, where $n_0$ is the initial excitation site. The discrete time evolution of the system can be monitored by considering the following two dynamical variables: the total optical power $P^{(m)}$ at time step $m$ in the superlattice, defined by 
\begin{equation}
P^{(m)}= \sum_n \left( |u_n^{(m)}|^2+|v_n^{(m)}|^2 \right)
\end{equation}
and the normalized second-order moment
\begin{equation}
M_2^{(m)}= \frac{\sum_{n} (n-n_0)^2 \left( |u_n^{(m)}|^2+|v_n^{(m)}|^2 \right)}{P^{(m)}}.
\end{equation} 
Such two dynamical variables can be readily measured in a photonic experiment, and they can provide useful information about spectral and localization properties of the superlattice \cite{e1d,e2,e5}. Since the dynamics is non-Hermitian, the total optical power $P^{(m)}$ is not conserved, for both $h<h_c$ and $h>h_c$. However, when the quasi-energy spectrum is entirely real, i.e. for $h<h_c$, the growth of $P^{(m)}$ with $m$ is not secular and $P^{(m)}$ remains limited as $m$ grows. Conversely, when $h>h_c$ the growth of $P^{(m)}$ with $m$ is unbounded. The localization properties of the superlattice are captured by the dynamical evolution of the second-order moment $M_2^{(m)}$. 
When all eigenfunctions are exponentially localized, $M_2^{(m)}$ remains bounded as $m$ increases, at a value of the order of the typical localization length of the eigenstates. Conversely, when the eigenfunctions are delocalized, the growth of $M_2^{(m)}$ is not bounded. Given the finite number of sites $M$ in the superlattice arranged in a ring geometry [Fig.1(b)], in our numerical simulations the presence of delocalized wave functions yields for $M_2^{(m)}$ a large (yet finite) value, of the order of $M$, in the delocalized phase.
The different types of phase transitions, displayed by the two models defined by the potential sequences given by Eqs.(19) and (22), are clearly illustrated in Figs.8-11. The figures show typical dynamical evolution of the photonic quantum walk on a superlattice under periodic boundary conditions for initial single-site excitation. Panels (b) in the figures show the behavior of the total optical power $P^{(m)}$, clearly indicating that both models show a spectral phase transition as $h$ is increased above $h_c$, signaled by the unbounded growth of the total optical power $P_n^{(m)}$. Panels (c) in Figs.8-11 show the corresponding behavior of the second-order moment. Note that only the first model, defined by the potential sequence Eq.(19), shows a localization phase transition, signaled by a marked different growth  of second-order moment as $h$ is increased above the critical value $h_c$ [compare Figs.8(c) and 9(c)]. More specifically, for $h<h_c$ (Fig.8) the excitation remains tightly localized close to the originally excited site, whereas for $h>h_c$ (Fig.9)  the excitation can spread over the entire lattice, with a characteristic unidirectional drift associated to the non-reciprocal hopping induced by the imaginary gauge field. Conversely, in the second model, defined by the potential sequence given by Eq.(22), the second-order moment can reach large values (of the order of $\sim M$) for an imaginary gauge field $h$ both below and above the critical value $h_c$. More specifically, for $h<h_c$ (Fig.10) the excitation can drift far away than its original position and can spread in one half of the lattice, i.e. between the two barriers, with the characteristic unidirectional drift associated to the non-reciprocal hopping. The only difference at $h>h_c$ (Fig.11) is that, owing to the very strong imaginary gauge field, the excitation can superpass the potential barrier and the excitation can drift over the entire lattice.}
\begin{figure}[htbp]
  \includegraphics[width=86mm]{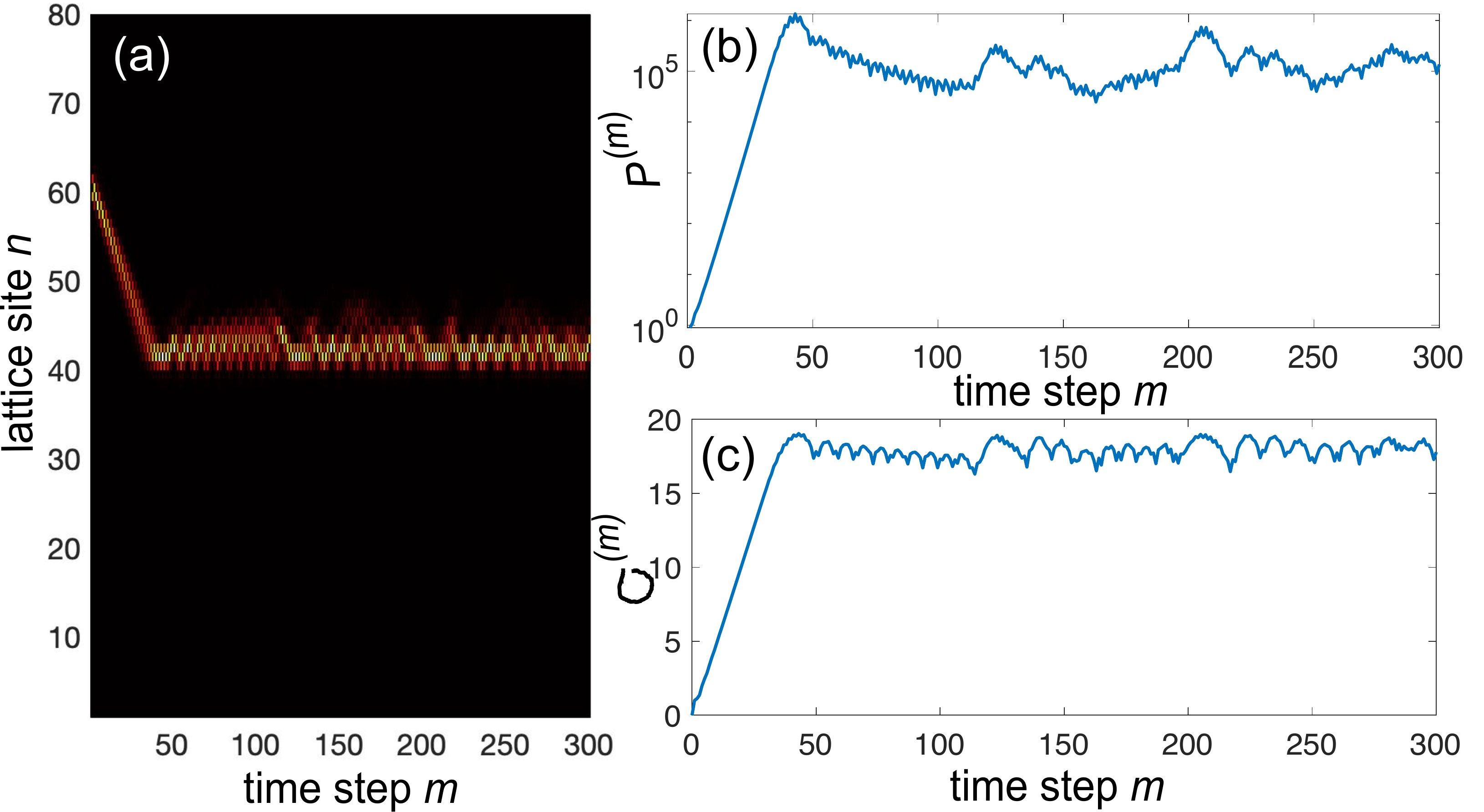}\\
  \caption{(color online)  Dynamical signatures of the phase transition in the photonic quantum walk model, defined by the potential sequence given by Eq.(22) (potential barrier model). Parameter values are as in Fig.7, with a number of sites $M=80$ and imaginary gauge field $h=0.45$, below the critical value $h_c=0.58$. The system is initially excited at the site $n=n_0=60$. (a) Numerically-computed behavior of the evolution of the normalized intensity distribution $(|u_n^{(m)}|^2+|v_n^{(m)}|^2)/P^{(m)}$ on a pseudo color map in the $(n,m)$ plane. (b) Behavior of the total optical power $P^{(m)}$ on a log scale. (c) Behavior of the square root of the normalized second-order moment $\sigma^{(m)}= \sqrt{M_2^{(m)}}$. }
\end{figure}
\begin{figure}[htbp]
  \includegraphics[width=86mm]{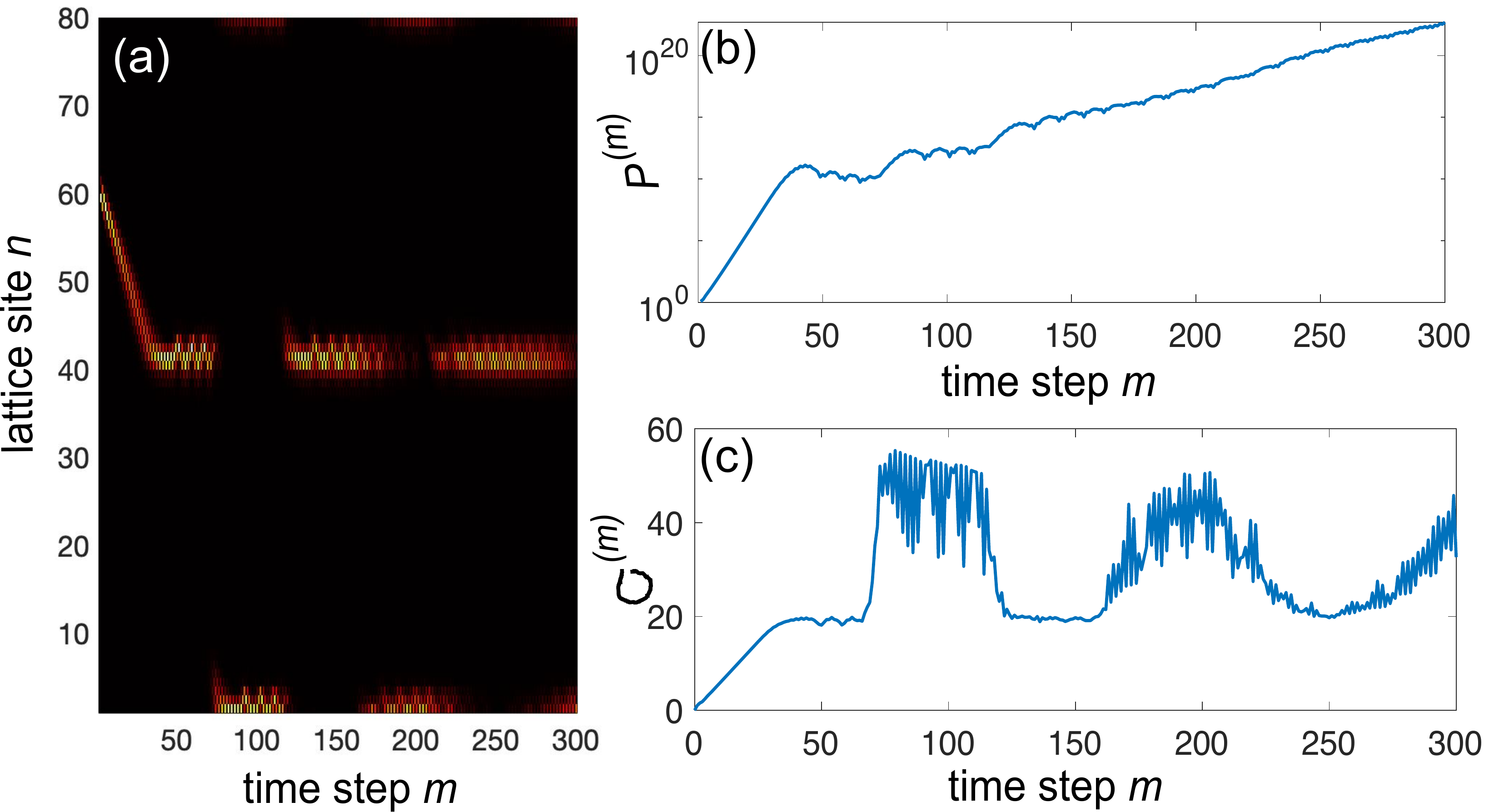}\\
  \caption{(color online) Same as Fig.10, but for $h=0.7$.}
\end{figure}
\section{Conclusion}
In this work we unravelled  spectral phase transitions in one-dimensional non-Hermitian superlattices subjected to an imaginary gauge field and possessing $M$ sites in each unit cell, in the large $M$ limnit. It has been demonstrated that in 
 models displaying nearly flat bands a smooth phase transition, from quasi entirely real to complex energies, can be observed as the imaginary gauge field is increased above a critical value, and that the phase transition becomes exact in the infinite $M$ limit. 
In superlattices with random or incommensurate disorder, the spectral phase transition is accompanied by a localization-delocalization transition of the eigenfunctions within each unit cell, a phenomenon which can be regarded, for a finite $M$, a precursor of the non-Hermitian delocalization transition originally predicted by Hatano and Nelson. However, the main result of the present work is that there exist superlattices where, in the large $M$ limit, the spectral phase transition {\em does not correspond} to a simultaneous localization/delocalization transition of the wave functions. The main reason thereof is that the spectral phase transition is observed in any superlattice model displaying miniband flattening in the large $M$ limit, with a bandwidth of each miniband exponentially vanishing with the unit cell size $M$. Wave function localization induced by disorder, like in the Anderson or Aubry-Andr\'e models, obviously corresponds to miniband flattening, when we approximate the disordered crystal with a superlattice with a large number of sites $M$ in the unit cell \cite{Th1}. In this case, the NH delocalization transition originally predicted by Hatano and Nelson is found: as the imaginary gauge field is increased, the emergence of a complex eigenenergy is associated to a delocalization of the corresponding wave function. However, band flattening does not necessarily require localization of the wave functions, as we showed in an illustrative example of an ordered superlattice corresponding to a sequence of potential barriers on a lattice. The predicted phenomena have been extended by considering non-Hermitian discrete-time photonic quantum walks, where synthetic superlattices with controllable potentials and imaginary gauge fields can be realized with existing experimental apparatus. Our results shed major insights into phase transitions in non-Hermitian lattices, indicating that the coincidence of spectral and metal-insulator phase transitions observed so far in non-Hermitian disordered systems is not a universal rule.

\acknowledgments
The author acknowledges the Spanish State Research
Agency through the Severo Ochoa and Maria de Maeztu Program for Centers and Units of Excellence in R \&D (Grant No. FQ 522 MDM-2017-0711).

\appendix
\section{Energy dispersion curves of the superlattice minibands: the near-flat band limit}
Let us consider the tight-binding superlattice in the Hermitian limit, i.e. for $h=0$, and let us indicate by $E_l(k,0)$ the dispersion curves of the $M$ minibands ($l=1,2,...,M$). In this Appendix we wish to provide an approximate form of the dispersion curves in the large $M$ limit when the superlattice displays near flat bands. To this aim, let us remind that the dispersion curves are obtained as the eigenvalues of the Bloch Hamiltonian
\begin{widetext}
\begin{equation}
\mathcal{H}(k)= \left(
\begin{array}{ccccccc}
V_1 & J & 0 & ...& 0 & 0& J \exp(-ikM) \\ 
J & V_2 & J & ... & 0 & 0& 0 \\
0 & J & V_3 & ... & 0 & 0 & 0 \\
... & ... & ... & ... & ... & ... & ... \\
0 & 0 & 0 & ... & J & V_{M-1} & J \\
J \exp(ikM) & 0 & 0 & ... & 0 & J & V_M
\end{array}
\right).
\end{equation}
\end{widetext}
Let us indicate by $\phi_n$ the $l$-th eigenfunction of $\mathcal{H}(k)$ at $k=0$, with eigenenergy $E(0)=E_l(k=0,0)$, and let us assume that $\phi_n$ is tightly confined near the center of the unit cell, with rapidly exponentially-decaying tails near the edges of the unit cell with a characteristic decay rate $\gamma_m$, as schematically shown in the bottom panels of Fig.3 of the main text. Note that, since for $k=0$ the matrix $\mathcal{H}(k)$ is Hermitian with real elements, without loss of generality we can take $\phi_n$ to be real as well. Let us assume the normalization condition $\sum_n \phi_n^2=1$. With such a normalization condition, the values $\phi_1$ and $\phi_M$ of the wave function at the edge sites of the unit cell are, in modulus, much smaller than the peak value of $| \phi_n|$ near the center of the unit cell, and they exponentially vanish with the size $M$ of the unit cell.  As a consequence, $\phi_n$ provides an approximate form the the eigenfunction of $\mathcal{H}(k)$ also for $k \neq 0$, since the eigenvalue equation becomes weakly sensitive to the precise values of the matrix elements $\mathcal{H}_{1,M}$ and $\mathcal{H}_{M,1}$. To calculate the form of the eigenfunctions and corresponding correction to the energy for $k \neq 0$, we can thus employ a perturbative analysis. Let us write the matrix $\mathcal{H}(k)$ in the form $\mathcal{H}(k)=\mathcal{H}_0+ \mathcal{H}_1(k)$, where 
\begin{widetext}
\begin{equation}
\mathcal{H}_0= \left(
\begin{array}{ccccccc}
V_1 & J & 0 & ...& 0 & 0& 0 \\ 
J & V_2 & J & ... & 0 & 0& 0 \\
0 & J & V_3 & ... & 0 & 0 & 0 \\
... & ... & ... & ... & ... & ... & ... \\
0 & 0 & 0 & ... & J & V_{M-1} & J \\
0 & 0 & 0 & ... & 0 & J & V_M
\end{array}
\right).
\end{equation}
and 
\begin{equation}
\mathcal{H}_1(k)= \left(
\begin{array}{ccccccc}
0 & 0 & 0 & ...& 0 & 0& J \exp(-ikM) \\ 
0 & 0 & 0 & ... & 0 & 0& 0 \\
0 & 0 & 0 & ... & 0 & 0 & 0 \\
... & ... & ... & ... & ... & ... & ... \\
0 & 0 & 0 & ... & 0 & 0 & 0 \\
J \exp(ikM) & 0 & 0 & ... & 0 & 0 & 0
\end{array}
\right).
\end{equation}
\end{widetext}
i.e. $\mathcal{H}_0$ is a tridiagonal matrix with real elements independent of $k$, whereas $\mathcal{H}_1(k)$ keeps the dependence on $k$ with only two non-vanishing elements $(\mathcal{H}_1)_{1,M}$ and $(\mathcal{H}_1)_{M,1}$. We can then solve the eigenvalue equation $(\mathcal{H}_0+\mathcal{H}_1(k) ) \phi_n(k)=E(k) \phi_n(k)$ perturbatively, considering the action of the matrix $\mathcal{H}_1(k)$ as a small perturbation, which is justified by the fact that the elements $\phi_1(k)$ and $\phi_M(k)$ are vanishing in the large $M$ limit. 
After letting $\phi_n(k)=\phi_n^{(0)}+\phi_n^{(1)}(k)+...$ and $E(k)=E^{(0)}+E^{(1)}(k)+...$, at leading order we have 
\[
\mathcal{H}_0 \phi_n^{.(0)}=E^{(0)} \phi_n^{(0)}
\]
which can be solved with the normalization condition $\langle \phi_n^{(0)} | \phi_n^{(0)} \rangle = \sum_n \left( \phi_n^{(0) } \right)^2=1$. We assume that the zeroth-order form of the wave function $\phi_n^{(0)}$, for a given miniband, corresponds to a  tightly confined function, near the center of the unit cell, with rapidly exponentially-decaying tails near the edges of the unit cell with a characteristic decay rate $\gamma_m$.  At first order one obtains
\[
\left( \mathcal{H}_0-E^{(0)} \right)  \phi_n^{(1)}=-\mathcal{H}_1(k) \phi_n^{(0)}+E^{(1)} (k) \phi_n^{(0)}
\]
which can be solved for $\phi_n^{(1)}$ provided that the solvability condition
\begin{equation}
E^{(1)}(k)= \langle \phi_n^{(0)} | \mathcal{H}_1(k) | \phi_n^{(0)} \rangle=2 J \cos (kM) \phi_1^{(0)} \phi_M^{(0)}
\end{equation}
is satisfied. Therefore, al leading order the dispersion curve of the given miniband can be written as
\begin{eqnarray}
E(k,0) &  \simeq & E^{(0)}+E^{(1)}(k) \nonumber \\
 & = & E(0,0)+\Delta [1- \cos (kM)]
\end{eqnarray}
where we have set $E(0,0)=E^{(0)}+2 J \phi_1^{(0)} \phi_M^{(0)}$ and
\begin{equation}
\Delta=-2 J \phi_1^{(0)} \phi_M^{(0)}.
\end{equation}
Finally, let us discuss the scaling of the miniband width $2 \Delta$ on the size $M$ of the unit cell, which entails to estimate the product $\phi_1^{(0)} \phi_M^{(0)}$, under the normalization condition $\sum_n \left( \phi_n^{(0)} \right)^2=1$. 
To this aim, let us indicate by $\rho$, with $0< \rho  \leq 1$, the fraction of the unit cell where the wave function $\phi_n^{(0)}$ displays exponential (evanescent) decay, with decay rate $\gamma_m$ independent of $M$, and by $(1-\rho)$ the fraction of the unit cell where the wave function is extended (oscillating). For example, for the incommensurate Aubry-Andr\'e model of Fig.3(a) we have $\rho=1$, whereas for the ordered superlattice with potential barriers of Fig.3(b) we have $\rho=1/2$. The dominant scaling of the product $\phi_1^{(0)} \phi_M^{(0)}$ with $M$ then reads
\[
\phi_1^{(0)} \phi_M ^{(0)} \sim \exp(-\sigma M)
\]
 with $\sigma= \rho \gamma_m$.

\section{Quasi-energy minibands of the photonic quantum walk superlattice}
Let us assume PBC and let us look for a solution to Eqs.(13a) and (13b) in the form of extended Bloch waves with quasi-energy $E=E(k,h)$, i.e. of the form
\begin{equation}
 u_n^{(m)}=U_n \exp[-iE(k,h) m] , \; \;   v_n^{(m)}=V_n \exp[-iE(k,h) m]
 \end{equation}
 where the amplitudes $U_n$, $V_n$ satisfy the boundary conditions
 \begin{equation}
 U_{n+M}=U_n \exp(ikM) \; , \;\; V_{n+M}=V_n \exp(ikM)
 \end{equation}
and where $k$ is the Bloch wave number, which varies in the interval $(-\pi/M, \pi/M)$. Substitution of Eq.(B1) into Eqs.(13a) and (13b) and taking onto account the boundary conditions  (B2), one readily obtains that the following eigenvalue equation should be satisfied
\begin{equation}
\exp[-iE(k,h)]
\left(
\begin{array}{c}
U \\
V
\end{array}
\right)= 
\left(
\begin{array}{cc}
\mathcal{A} & \mathcal{B}\\
\mathcal{C} & \mathcal{D}
\end{array}
\right) \left(
\begin{array}{c}
U \\
V
\end{array}
\right) 
\end{equation}
for the amplitudes $U \equiv (U_1,.U_2,...,U_M)^T$ and $V \equiv (U_1,.U_2,...,U_M)^T$. In Eq.(B3), the four $M \times M$ matrices $\mathcal{A}$, $\mathcal{B}$, $\mathcal{C}$, and $\mathcal{D}$ are defined by
\begin{eqnarray}
\mathcal{A}= \exp(h) \cos \beta \Phi \Theta \;, \;\;  \mathcal{B}=i \exp(h) \sin \beta \Phi \Theta \nonumber \\
\mathcal{C}= i \exp(-h) \sin \beta \Phi \Theta^{\dag} \;, \;\; \mathcal{D}= \exp(-h) \cos \beta  \Phi \Theta^{\dag} \nonumber
\end{eqnarray}
 where $\Phi$ is the $M \times M$ diagonal matrix with the potential terms $\exp(i \varphi_n)$ on the main diagonal, i.e.
 \begin{equation}
 \Phi_{n,m}= \exp(i \varphi_n) \delta_{n,m}
 \end{equation}
 and
 \begin{equation}
 \Theta_{n,m}= \delta_{n,m-1}+ \exp(ikM) \delta_{n,M}\delta_{m,1}.
 \end{equation}
 ($n,m=1,2,...,M$).  The quasi energy dispersion curves $E=E_l(k,h)$ of the $2M$ minibands of the superlattice are finally obtained by solving the determinantal equation
 \begin{equation}
 \left| \exp[-i E(k,h)]- 
 \left(
\begin{array}{cc}
\mathcal{A} & \mathcal{B}\\
\mathcal{C} & \mathcal{D}
\end{array}
\right)
\right|=0.
 \end{equation}
 It can be readily shown that $E_l(k,h)=E_l(k-ih,0)$, i.e. the dispersion curves of the NH quantum walk are obtained from the ones of the Hermitian case after complexification of the Bloch wave number via the substitution $k \rightarrow k-ih$.

\end{document}